\documentstyle[aps,preprint]{revtex}
\def\gsim
 {\raise2pt\hbox
 {$\displaystyle{}\mathrel{\mathop>_{\raise1pt\hbox{$\sim$}}}{}$}}
\def\lsim
 {\raise2pt\hbox
{$\displaystyle{}\mathrel{\mathop<_{\raise1pt\hbox{$\sim$}}}{}$}}
\draft
\begin{document}
\title{Macroscopic Quantum Tunneling and Dissipation of Domain Wall
in Ferromagnetic Metals}
\author{Gen Tatara}
\address{ The Institute of Physical and Chemical Research (RIKEN),
Wako, Saitama 351-01, Japan}
\author{Hidetoshi Fukuyama}
\address{ Department of Physics, University of Tokyo,
        7-3-1 Hongo, Tokyo 113, Japan }
\date{\today}
\maketitle
\newcommand{\beq}{\begin{equation}}
\newcommand{\eeq}{\end{equation}}
\newcommand{\beqa}{\begin{eqnarray}}
\newcommand{\eeqa}{\end{eqnarray}}
\newcommand{\sigbf}{\mbox{\boldmath$\sigma$}}
\newcommand{\Mv}{{\bf M}}
\newcommand{\nv}{{\bf n}}
\newcommand{\Jv}{{\bf J}}
\newcommand{\kv}{{\bf k}}
\newcommand{\qv}{{\bf q}}
\newcommand{\xv}{{\bf x}}
\newcommand{\half}{\frac{1}{2}}
\newcommand{\kf}{k_{\rm F}}
\newcommand{\kfu}{k_{{\rm F}\uparrow}}
\newcommand{\kfd}{k_{{\rm F}\downarrow}}
\newcommand{\kfs}{k_{{\rm F}\sigma}}
\newcommand{\ef}{\epsilon_{\rm F}}
\newcommand{\ek}{\epsilon_{\bf k}}
\newcommand{\efs}{\epsilon_{{\rm F}\sigma}}
\newcommand{\eks}{\epsilon_{{\bf k}\sigma}}
\newcommand{\ekqs}{\epsilon_{\kv+\qv,\sigma}}
\newcommand{\ekd}{\epsilon_{\kv\downarrow}}
\newcommand{\ekqu}{\epsilon_{\kv+\qv,\uparrow}}
\newcommand{\fks}{f_{\kv\sigma}}
\newcommand{\fkqs}{f_{\kv+\qv,\sigma}}
\newcommand{\fkd}{f_{\kv\downarrow}}
\newcommand{\fkqu}{f_{\kv+\qv,\uparrow}}
\newcommand{\cel}{c_{\xv\sigma}}
\newcommand{\ael}{a_{\xv\sigma}}
\newcommand{\cd}{c_{\xv\sigma}^\dagger}
\newcommand{\ad}{a_{\xv\sigma}^\dagger}
\newcommand{\thx}{\theta_{\xv}}
\newcommand{\phx}{\phi_{\xv}}
\newcommand{\Hu}{H_{\rm U}}
\newcommand{\Hus}{H_{\rm U}^{(\rm slow)}}
\newcommand{\Huf}{H_{\rm U}^{(\rm fast)}}
\newcommand{\Ms}{M_0}
\newcommand{\Cpm}{C_{+-}}
\newcommand{\Csig}{C_{\sigma}}
\newcommand{\oml}{\omega_\ell}
\newcommand{\chiu}{\chi^0_{\uparrow}}
\newcommand{\chid}{\chi^0_{\downarrow}}
\newcommand{\DSone}{\Delta S^{(1)}}
\newcommand{\DStwo}{\Delta S^{(2)}}
\newcommand{\nzero}{n_0}
\newcommand{\Awall}{A_{\rm w}}
\newcommand{\Spin}{S}
\newcommand{\ktil}{\tilde k_0}
\newcommand{\Vtil}{\tilde V}
\newcommand{\lamtil}{\tilde \lambda}
\newcommand{\mels}{m'}
\newcommand{\nels}{n'}
\begin{abstract}
The depinning of a domain wall in ferromagentic metal via macroscopic quantum
tunneling is studied based on the Hubbard model.
The dynamics of the magnetization verctor is shown to be governed by
an effective action of Heisenberg model with a term non-local in time that
describes the dissipation due to the conduction electron.
Due to the existence of the Fermi surface there exists Ohmic dissipation even
at zero
temperature, which is crucially different from the case of the insulator.
Taking into account the effect of pinning and the external magnetic field the
action is rewritten in terms of a collective coordinate, the position of
the wall, $Q$.
The tunneling rate for $Q$ is calculated by use of the instanton method.
It is found that the reduction of the tunneling rate due to the dissipation
is very large for a thin domain wall
with thickness of a few times the lattice spacing, but is negligible for
a thick domain wall.
Dissipation due to eddy current is shown to be negligible for a wall
of mesoscopic size.

\end{abstract}
\section{Introduction}
\label{SECint}

Domain wall in a ferromagnet is a soliton connecting two stable spin
configurations separated by the energy barrier due to the anisotropy
(Fig. \ref{FIGDW}).
It is generally pinned by defects where the spins are easier to rotate.
\begin{figure}[bt]
\vspace{3cm}
\caption{ A  configuration of magnetization of a planar domain wall (a)
where the spin configuration spatially varies in $x$-direction, which is
 perpendicular to the easy plane of the spin.
 A planer wall with the easy axis perpendicular to the wall is shown in
(b)\protect\cite{dwb}.
  \label{FIGDW} }
\end{figure}
The wall gets depinned by the external magnetic
field (Fig. \ref{FIGDWMQT}). This depinning process involves the motion
of an extended object, but it can be simplified if one neglects the bending
of the wall and describes the process as a motion of a collective varialbe,
the position of the wall, $Q$. The potential for $Q$ created by defects (
Fig. \ref{FIGDWpot}(a)) is deformed by the external magnetic field $H$,
and the position of the wall at the
pinning center becomes metastable (Fig. \ref{FIGDWpot}(b)).
The barrier height is reduced as the field is increased and finally
vanishes at the coercive field, $H_c$.
At high temperature the depinning occurs by the thermal fluctuation of the
variable $Q$ overcoming the barrier, and at low temperature it is due to the
quantum tunneling of $Q$.
Since the wall contains a large number of spins it is a macroscopic
quantum tunneling (MQT).

This MQT of domain wall was investigated theoretically by
Stamp\cite{Sta,CIS,SCB}.
His consideration on an insulating magnet in three dimensions is based on the
ferromagnetic Heisenberg model with a uniaxial anisotropy.
By use of a classical solution of planar domain wall, the
hamiltonian describing the position of the  wall is obtained.
The pinning potential by a defect is assumed to be of a short range.
The experiments of the tunneling of the wall from the metastable minima are
carried out in a magnetic field close to this coercive field ($H<H_c$) for the
rate to be large enough to be observed.
Stamp obtained the expression of the tunneling rate
as functions of the coercive
field, saturation magnetization, the number of spins in the wall and the
external field, and concluded that
depinning of a wall due to MQT is observable  even for a large wall
containing about $10^{10}$ spins.
For observing the MQT,  highly anisotropic
materials (such as those containing rare-earth) would be suitable, since the
wall will be small (thin) in these materials and also the crossover temperature
from the thermal to the quantum regime can be higher.
\begin{figure}[tb]
\vspace{2cm}
\caption{(a):A domain wall pinnied at a defect. (b):Depinning of the wall
 by the external magnetic field.  \label{FIGDWMQT}  }
\vspace{3.3cm}
\caption{The potential for the domain wall in terms of $Q$, the position of the
center of the wall; (a) without the
 magnetic field and (b) with the magnetic field.  \label{FIGDWpot}  }
\end{figure}

In macroscopic quantum systems, the coupling to the environment, which results
in dissipation, may become significant, and the effect needs to be estimated.
In their seminal paper Caldeira and Leggett\cite{CL} presented a formalism
that can incorporate
the dissipation in quantum mechanics by use of the imaginary time path
integral.
There the dissipation is expressed by the action non-local in time.
Based on this formulation, Caldeira and Leggett investigated the quantum
tunneling of a macroscopic variable, and found that the dissipation
generally reduces the tunneling rate.

As sources of dissipation in the MQT of domain wall, Stamp considered the
magnon (spin wave) and phonon.
Because of the gap due to the anisotropy, the Ohmic dissipation
due to the spin wave vanishes at absolute zero  and is quantitatively
negligible at low temperature.
The phonon was also shown to be irrelevant.
The effects of conduction electrons were briefly touched by Chudnovsky
{\it et al.}\cite{CIS} on a phenomenological ground.
Concerning the dissipation in single domain magnets, the magnetoelastic
coupling to
phonons was considered, but the effect turned out be negligible in actual
situations\cite{GK}.
The effects of spin wave have not been discussed, but are expected to be
negligible since this degree of freedom
is frozen out at very low temperature due to the anisotropy gap.
Recently it was claimed that the dissipation due to hyperfine coupling to the
nuclear spin is significant\cite{Gargnuc}.

The depinning of the domain wall is observed by measuring the relaxation
of the magnetization, and MQT is indicated by the relaxation rate going
to a finite constant value as $T\rightarrow 0$.
The first indication of MQT was obtained in a bulk ferromagnet
of SmCo$_{3.5}$Cu$_{1.5}$\cite{UB}, but detailed study
 was carried out on small ferromagnetic particles of
Tb$_{0.5}$Ce$_{0.5}$Fe$_2$ of diameter of about 150\AA\cite{PSB}.
Small particles are suitable for such experiments since in the bulk
sample many walls with different size participate in the relaxation
process.
The temperature independent relaxation was observed for $T\lsim0.6$K, and
the result was claimed to be consistent with theory\cite{Sta} of MQT of
a domain wall in a ferromagnet without dissipation.

Although the materials employed in these experiments are generally matallic,
there has been
no theoretical work paying full attention to this fact.
In metals the behavior of the magnetic object may not be so simple as in
insulators.
Actually the magnetization arises due to the polarization of the electron
and at the same time these electrons have
dissipative effects for the magnetization vector.
To take into account these points  in a systematic way our analysis\cite{TF}
of the MQT of a domain wall is based on the Hubbard model.
We consider the case of absolute zero, $T=0$ ,
since we are interested only in the quantum tunneling present at low
temperature.
The calculation is carried out in the continum\cite{continum}.
The magnetization vector is expressed as an expectation value
of the electron spin operator.
For the treatment of a slowly varying field like a domain wall, the locally
 rotated frame for the electron is convenient.
By use of this frame, the effective action that describes
the low energy dynamics of magnetization vector is obtained
by integrating out the electron degrees of freedom
in the imaginary time path integral.
As far as low energy behavior is concerned,
the modulation of the magnitude of the magnetization is irrelevant, and
only its direction can be a dynamical variable.
Then the part of the effective action that is local in time ({\it i.e.,}
instantaneous) has the same form as that of  a ferromagnetic Heisenberg
model; {\it i.e.,} there is no formal difference from the case of the
insulator but the parameters have different physical origin.
On the other hand, the non-local ({\it i.e.,} retarded) part of the
effective action, which describes the dissipative effect of itinerant
electron on the motion
of the magnetization, is  crucially different from the case of an insulator.
Due to the Stoner excitaion, which is the gapless excitation of spin flip
across the fermi surface, the Ohmic dissipation is present even at zero
temperature.
It is shown that this dissipation reduces the tunneling rate very much
particularly for a thin
domain wall of thickness comparable to the inverse of the difference
of the fermi momenta $(\kfu-\kfd)^{-1}$.
In the case of strong ferromagnet, on the other hand, dissipation resulting
from the Stoner excitation is reduced. In such a case the existence of
non-magnetic band in actual systems is to be taken into account.
In contrast to these effects from the Stoner excitation, the effects of Ohmic
dissipation due to charge current (eddy current) will be found to be negligible
in a meso- or microscopic wall we are interested in.

This paper is organized as follows.
\S\ref{SECeff} is devoted to the derivation of the effective action for
the magnetization vector on the basis of the Hubbard model.
In \S\ref{SECdw} the MQT of a domain wall is studied based on this action.
The expression of the tunneling rate including the
dissipative effect from itinerant electron is obtained there.
The dissipative effects resulting from the non-magnetic band are calculated
in
\S\ref{SECsd}, and in \S\ref{SECeddy} the  effects due to the eddy current
are estimated.
Discussions and conclusion are given in \S\ref{SECdisc}.

\section{Derivation of an Effective Action}
\label{SECeff}
\subsection{Model}
We consider the Hubbard model
in the imagnary time path integral whose Lagrangean is given by\cite{magfield}
\begin{equation}
L=\sum_\xv \sum_{\sigma} \left( \cd (\partial_\tau-\epsilon_{\rm F})
\cel+\frac{1}{2m} |\nabla \cel|^2 \right)
   + U \sum_\xv  n_{\xv\uparrow}n_{\xv\downarrow} , 		\label{L0}
\end{equation}
where $\cel$ is an electron operator at site $\xv$ with spin $\sigma(=\pm)$
and $n_{\xv\sigma}\equiv\cd\cel$,
$\epsilon_{\rm F}$ being the chemical potential.
For simplicity, a parabolic dispersion with mass $m$ is assumed for band
energy.
The ferromagnetism is induced by the Coulomb repulsion ($U$) term,
which is rewritten by introducing the magnetization vector $\Mv$
 by use of the Hubbard-Stratonovich transformation.
The vector is expressed as $\Mv(\xv,\tau)\equiv M(\xv,\tau) \nv(\xv,\tau)$,
where $\nv$ is a slowly varying unit vector that represents the direction of
the magnetiazation and the magnitude of the magnetization is given by
\beq
M(\xv,\tau)\equiv <(c^\dagger {\bf \sigbf} c) >_\xv \cdot \nv(\xv,\tau).
\label{Mdef}
\eeq
The Lagrangian is then written as
\beq
L=\sum_{\kv\sigma} c^\dagger_{\kv \sigma}(\partial_\tau + \epsilon_\kv)
c_{\kv\sigma}
  -U\sum_\xv\Mv(\xv) (c^\dagger {\bf \sigbf} c)_\xv
 +\frac{U}{2} \sum_\xv M(\xv)^2.
  \label{L1}
\eeq

The spatial variation of $\nv(\xv)$ accompanying with a
domain wall is much slower compared to
the inverse fermi momentum of the electron $\kf^{-1}$.
For the analysis of such a slowly varying field,
the local frame of electron\cite{Pra} is convenient for the perturbative
treatment
of the interaction between the domain wall and the conduction electron
to be valid. In this frame the $z$-axis
of the electron is chosen in the direction of the local magnetization vector
$\nv(\xv)$.
The electron operator in the new frame $\ael$ is related to $\cel$ in the
origial one as
\beq
  a_{\xv\sigma} = \sigma\cos{\frac{\theta}{2}}c_{\xv,\sigma}
               + e^{-i\sigma\phi}\sin{\frac{\theta}{2}} c_{\xv,-\sigma}
\label{loc}
\eeq
where $(\theta,\phi)$ are the polar coordinates of vector
$\nv(\xv,\tau)$.
We neglect in the following calculation the spatial variation of the
magnitude $M(\xv)$ of the magnetization ({\it i.e.,} $M(\xv)\equiv M$), since
the fluctuation of $M(\xv)$ has a finite mass.
Thus the relevant degree of freedom at low energy is only
the variation of the angle, ($\theta,\phi$), of the magnetization.
The $\ael$ electron is then polarised uniformly in the $z$-direction,
and it interacts with space-time variation of the magnetization
vector.
The interaction term arises from the kinetic energy term
$c^\dagger \dot c + |\nabla c|^2/(2m)$, and is written as
\beqa
\lefteqn{ H_{\rm int}=\sum_\xv \left[ \frac{i}{2}\dot{\phi}(1-\cos\theta)
 (a^\dagger\sigma_z a)
 \right. +\sum_{\pm} \half(\pm\dot{\theta}-i\sin\theta\dot{\phi})e^{\mp i\phi}
(a^\dagger\sigma_\pm a)
 }   \nonumber\\
&& +\frac{1}{4m}\left(\half (\nabla\theta)^2+(1-\cos\theta)(\nabla\phi)^2
\right)
    (a^\dagger a)  \nonumber\\
&& +\frac{1}{2}(1-\cos\theta)\nabla\phi \Jv_z(\xv)
   \left. +\frac{i}{2}\sum_{\pm}e^{\mp i\phi}
  (\mp\nabla\theta+i\sin\theta\nabla\phi) \Jv_\pm(\xv) \right],
\label{HI}
\eeqa
where $\Jv_\alpha (\xv)$ $(\alpha=\pm,z)$ are the spin currents of the
electron;
\beq
\Jv_\alpha (\xv,\tau)\equiv -\frac{i}{2m}[(a^\dagger\sigma_\alpha \nabla a)
- (\nabla a^\dagger\sigma_\alpha a)] ,
\label{curre}
\eeq
with $\sigma_\pm\equiv\sigma_x\pm i\sigma_y$.

The total Lagrangean written in the local frame is therefore given by
\beq
L=\sum_{\kv\sigma} a^\dagger_{\kv\sigma}(\partial_\tau+\epsilon_{\kv\sigma})
  a_{\kv\sigma}
   + \frac{U}{2}\sum_\xv M^2 +H_{\rm int} ,
  \label{L2}
\eeq
where the electron energy is now spin dependent,
$\epsilon_{{\kv}\pm}\equiv ({{\kv}^2}/{2m})\mp UM-\epsilon_{\rm F}$.

\subsection{Effective action}
The effective action of
the magnetization is obtained by integrating out the electron degree of
freedom treating the interaction term $H_{\rm int}$ perturbatively.
This is reasonable since the variation of the magnetization vector is
generally very slow compared to that of electron.
For the case of a domain wall with thickness $\lambda$,
this perturbative treatment corresponds to the expansion in terms of
$(\kfu\lambda)^{-1}$, $\kfu$ being the fermi momentum of the majority spin.
The effective action for the magnetization is obtained as
\beq
S_{\rm eff}(\theta,\phi,M)=S_{\rm MF}(M)+\Delta S(\theta,\phi,M),
\label{Seff}
\eeq
where $S_{\rm MF}$ is the well-known mean field action of ferromagnet;
$
 S_{\rm MF}\equiv
-{\rm tr}\ln(\partial_\tau+\epsilon_{\kv\sigma})+\beta\sum_\xv({U}/{2})
M^2 $,
and $\Delta S\equiv \int d\tau <H_{\rm int}>
-(1/2) \int d\tau\int d\tau' <H_{\rm int}(\tau)H_{\rm int}(\tau') >$.
The brackets indicate the expectation values with respect to the electron.

In further calculations, the variable $M$ is approximated by its mean
field value $\Ms$ determined by the stationary condition of
$\delta S_{\rm MF}(\Ms)/\delta \Ms=0$, {\it i.e.,}
\beq
\Ms=\frac{(2m\epsilon_{\rm F})^{\frac{3}{2}}a^3}{6\pi^2}
\left\{\left(1+\frac{U}{\epsilon_{\rm F}}\Ms\right)^{\frac{3}{2}}
 -\left(1-\frac{U}{\epsilon_{\rm F}}\Ms\right)^{\frac{3}{2}}
 \right\} . \label{MF}
\eeq
The fluctuation $\delta M$ of $M$ around $\Ms$ can be neglected for the
study of low energy behavior.
In terms of $\Ms$ the fermi momenta of electrons with spin up and down are
given as
$k_{{\rm F}\pm}\equiv (2m \epsilon_{\rm F})^{1/2}\left
(1\pm ({U}/{\epsilon_{\rm F})}\Ms\right)^{1/2}$.
The behaviors of quantities $\Ms$ and $(\kfs a)$
as a function of
$(U/\epsilon_{\rm F})$ (with the electron number per site, $n$, being
fixed to be unity) are shown in Fig. \ref{FIGMF}.
\begin{figure}[tb]
\vspace{2.7cm}
\caption{
  Mean field solutions of quantities
$\Ms$, $\tilde \kfs\equiv (\kfs a)$,
$\tilde J\equiv (J\Ms^2 a/\epsilon_{\rm F})$,
 and $\delta\equiv(\kfu-\kfd)/
(\kfu+\kfd)$ as a function of $\tilde U\equiv (U/\epsilon_{\rm F})$.
The ferromagnetism appears for $\tilde U\geq(2/3) $ and the complete
ferromagnetism is realized for $\tilde U\geq1$, where $\kfd$
vanishes\protect\cite{meanfield}.
  \label{FIGMF}}
\end{figure}

The dynamics of $(\theta,\phi)$ is determined by $\Delta S$, which up to
the order of $\nabla^2$ and $\partial_\tau$ is given as
\beqa
\lefteqn{  \Delta S
= \int d\tau \sum_\xv \left[ i\frac{M}{2}\dot{\phi}
(1-\cos\theta)
 +\frac{n}{4m}\left(\half (\nabla\theta)^2+(1-\cos\theta)(\nabla\phi)^2
 \right) \right]    }\nonumber\\
&& -\frac{1}{8}\int\! d\tau \! \int \! d\tau'\sum_{\xv\xv'} \sum_{ij} \left\{
 \left[(1-\cos\theta)\nabla_i\phi \right]
\left[(1-\cos\theta')\nabla_j\phi' \right]
 <\Jv_z^i(\tau,\xv) \Jv_z^j(\tau',\xv')>
 \right. \nonumber\\
&&
  + \left.  2 e^{- i(\phi-\phi')}
    (\nabla_i\theta-i\sin\theta\nabla_i\phi)
    (\nabla_j\theta'+i\sin\theta'\nabla_j\phi')
  <\Jv_+^i(\tau,\xv) \Jv_-^j(\tau',\xv')>
  \right\},  \nonumber\\
&& \label{DS}
\eeqa
where $n\equiv <(a^\dagger a)_\xv>$ is the electron number per site
and $(\theta',\phi')\equiv(\theta(\xv',\tau'),\phi(\xv',\tau'))$.

The correlation functions are calculated by the random phase approximation
(RPA) to the Coulomb interaction\cite{RPA}.
The Fourier transform of $<\Jv_+\Jv_->$ with  the thermal frequency
$\oml\equiv 2\pi\ell/\beta$ is evaluated, as
depicted in Fig. \ref{FIGRPA} diagramatically, to be
\beq
   <\Jv_+^i(\qv,\ell) \Jv_-^j(-\qv,-\ell)>  = \Cpm^{ij}
+\Cpm^{i}\Cpm^{j}\frac{2U}{1-2U\chi^0_{+-}} ,
\label{JJpm}
\eeq
where
\beqa
\Cpm^{ij}(\qv,\ell)&\equiv& \frac{1}{\Vtil}\frac{1}{4m^2}
 \sum_{\kv}(2k+q)^i(2k+q)^j\frac{\fkd-\fkqu}{\ekqu-\ekd-i\oml} \nonumber\\
\Cpm^{i}(\qv,\ell)&\equiv& \frac{1}{\Vtil}\frac{1}{2m}
 \sum_{\kv}(2k+q)^i\frac{\fkd-\fkqu}{\ekqu-\ekd-i\oml} ,\label{c+-}
\eeqa
and the irreducible spin susceptibility $\chi^0_{+-}$ is given by
\beq
\chi^0_{+-}(\qv,\ell)\equiv\frac{1}{\Vtil}\sum_{\kv}\frac{\fkd-\fkqu}
{\ekqu-\ekd-i\oml}  .\label{chi+-}
\eeq
Here $f_{\kv\sigma}$ is the fermi distribution function
$
f_{\kv\sigma}\equiv {1}/[{e^{\beta({\kv^2}/{2m}-\efs)}+1}],
$
with the fermi energy $\efs\equiv \epsilon_{\rm F}\pm U\Ms$.
The $z$-component $<J_zJ_z>$ is calculated similarly, but for our purpose
only its irreducible part is needed, which is given as
\beq
<\Jv_z^i(\qv,\ell) \Jv_z^j(-\qv,-\ell)>\simeq \frac{1}{\Vtil}\frac{1}{4m^2}
 \sum_{\kv\sigma}(2k+q)^i(2k+q)^j\frac{\fks-\fkqs}{\ekqs-\eks-i\oml}.
\eeq

\begin{figure}[tb]
\vspace{1.2cm}
\caption{
  The diagramatic expression of the spin current correlation function
$<\Jv_+\Jv_->$  in RPA.
  \label{FIGRPA} }
\end{figure}


\subsection{Effective Heisenberg model (local part)}
To make clear the meaning of the effective action, we devide the action
into two parts, that is local and non-local in time;
$\Delta S\equiv \Delta S_{\rm loc}+\Delta S_{\rm dis}$.
The local part up to order $\partial_\tau$ and $\nabla^2$ is calculated from
the $\omega_\ell=0, q=0$ components
of the correlation functions, which are obtained as
\beqa
\lim_{q\rightarrow0}<\Jv_+^i(\qv,0) \Jv_-^j(-\qv,0)>& =&
       \frac{(\kfu^{5}-\kfd^5)a^3}{60\pi^2m^2U\Ms} \delta_{ij}   \nonumber\\
\lim_{q\rightarrow0}<\Jv_z^i(\qv,0) \Jv_z^j(-\qv,0)>& =&
    \frac{n}{m}\delta_{ij}.
\eeqa
The local part
of the effective action results in\cite{TF}
\beq
\Delta S_{\rm loc}=\int d\tau \int {d^3\xv}
  \left[ i\frac{S}{a^3}\dot{\phi}(1-\cos\theta)+
   \frac{JS^2}{2}\left( (\nabla\theta)^2+\sin^2\theta(\nabla\phi)^2 \right)
  \right], \label{Sloc}
\eeq
where the spin is $S\equiv \Ms /2$ and the exchange coupling or the spin
stiffness $J$ ([J/m] in MKSA unit) is expressed by the parameters of
the itinerant electron as
\beq
J\equiv\frac{n}{ma^3\Ms^2}\left[1-\frac{(\kfu^{5}-\kfd^5)a^3}{30\pi^2 mnU\Ms}
\right] . \label{Jdef}
\eeq
The action Eq.(\ref{Sloc}) is of the same form as that of the ferromagnetic
Heisenberg model with $S=\Ms/2 $, as has already been
indicated\cite{Pra,RS}.
Hence, there is no formal difference from the case of an
insulator in the local part of the effective action, but
$S$ in the present itinerant system is a continuous variable in contrast
to the Heisenberg model.
The behavior of the exchange energy $(J\Ms^2 a/\epsilon_{\rm F})$ is
plotted as a function of
$(U/\epsilon_{\rm F})$ in Fig. \ref{FIGMF}.

\subsection{Dissipative part (non-local part)}
We now go on to the non-local part $\Delta S_{\rm dis}$. This term comes from
the second order contribution of $H_{\rm int}$ and it is this term that
contains the characteristic feature of the itinerant model, the dissipative
effect from the conduction electrons.
In the following analysis, we consider a planar wall with a spin
configuration
changing only in $x$-direction in space, and the spins lying in the
$yz$-plane; {\em i.e.,} $\phi=\pi/2$ (see Fig. \ref{FIGDW}(a)).
In this case, the dissipation arises only from $<\Jv_+ \Jv_->$ term, and
then the non-local part is given by
\beq
 \Delta S_{\rm dis}= \frac{1}{4}\int\! d\tau \! \int \! d\tau'
\frac{1}{\beta}\sum_\ell
 e^{i\oml(\tau-\tau')}
 \frac{1}{\Vtil}\sum_{\qv} |\Theta_\qv(\tau)-\Theta_\qv(\tau')|^2
     <\Jv_+^1(\qv,\ell) \Jv_-^1(-\qv,-\ell)> ,\label{DSexamp}
\eeq
where
$ \Theta_\qv(\tau)\equiv \sum_\xv
 e^{-i\qv\xv}\nabla\theta$.
To estimate this effect that dominates at low energy, the
analytical continuation
 to the real frequency $i\oml\rightarrow\omega\pm i0$ is
convenient\cite{ATF,TF}.
The important contribution is that of the Ohmic dissipation,  which is due to
the $\omega$-linear term in the imaginary part of the analytically continued
correlation function,
Im$<\Jv(\qv) \Jv(-\qv)>|_{i\omega_\ell\equiv\omega+i0}$
as $\omega\rightarrow0$\cite{superohmic}.
The Ohmic part of $\Delta S_{\rm dis}$ turns out to be\cite{TF}
\beq
 \Delta S_{\rm dis}= \frac{1}{4}\int\! d\tau \! \int \! d\tau'
 \frac{1}{\Vtil}\sum_{\qv}
\frac{|\Theta_\qv^i(\tau)-\Theta_\qv^i(\tau')|^2}{(\tau-\tau')^2}
  \lim_{\omega\rightarrow0}
\left(\frac{{\rm Im}<\Jv_+^1(\qv)
\Jv_-^1(-\qv)>|_{\omega+i0}}{\pi\omega}\right).
\label{ohm}
\eeq

After straightfoward algebra, the Ohmic contribution to the
imaginary part is calculated as\cite{TF}
\beq
{\rm Im}<\Jv^1_+(\qv) \Jv^1_-(-\qv)>|_{\omega+i0}=
  \pi\omega\frac{(\kfu^2-\kfd^2)^2}{4\pi^2|q|^3}\theta_{\rm st}(q) +
  O(\omega^3).  \label{Im}
\eeq
where the function $\theta_{\rm st}(q)$,
\beqa
\theta_{\rm st}(q)\equiv \left\{ \begin{array}{ccc} 1  & \mbox{  }  &
  (\kfu-\kfd)<|q|<(\kfu+\kfd)   \\
                                            0  &   &
  {\rm otherwise} \end{array}, \right.\label{st}
\eeqa
indicates that the Ohmic dissipation is due to the Stoner excitation,
which is a gapless excitation that flips a spin of an electron at the
fermi surface.
The effect of dissipation at zero temperature is finally given as
\beq
 \Delta S_{\rm dis}=
\frac{(\kfu^2-\kfd^2)^2 }{32\pi^2 }\int\frac{d^3q}{(2\pi)^3}
 \frac{1}{|q|^3} \theta_{\rm st}(q)  \int d\tau \int d\tau'
  \frac{ |\Theta_\qv^1(\tau)- \Theta_\qv^1(\tau')|^2}{(\tau-\tau')^2}
.  \label{Sdis}
\eeq
The total action for the direction of the magnetisation vector is
$\Delta S_{\rm loc}+\Delta S_{\rm dis}$.
Since $\Delta S_{\rm dis}$ is positive definite, the tunneling rate is
seen to be always reduced by this dissipation effect within the present
semi-classical approximation.
In the case of a complete ferromagnet, $(U/\epsilon_{\rm F})\geq1$,
$\kfd=0$ and then $\Delta S_{\rm dis}$ is vanishing.

\section{Effects of Dissipation on MQT of a Domain Wall}
\label{SECdw}
In the previous section, we have derived based on the Hubbard model an
effective action $\Delta S_{\rm loc}+\Delta S_{\rm dis}$
describing the slowly varying direction of the magnetization vector.
By use of  this effective action, we will investigate the quantum tunneling
of a domain wall in the case as shown in Fig.\ref{FIGDW}(a).
To do this, we need first to describe the domain wall motion in terms of
the tunneling variable, the coordinate of the the wall $Q$.
The dissipative effect $\Delta S_{\rm dis}$ is to be estimated by
use of the variable $Q$.

\subsection{MQT of a domain wall}
\label{SECdw1}
Let us first study the local part of the action $\Delta S_{\rm loc}$.
Without $\Delta S_{\rm dis}$, the analysis goes the same as in the
insulator case\cite{Sta,CIS,SCB}.
For the domain wall to exist, anisotropy energy is needed, which we
 shall add phenomenologically to the action $\Delta S_{\rm loc}$ as the
 energy $-K$[J/m$^3$] in the $z$-direction as an easy axis and
$+K_{\perp}$ in the $x$-direction as a hard axis\cite{demag,kperp,aniel}.
The action we deal with is therefore given by the following
 (including $\Delta S_{\rm dis}$)
\beqa
 S(\theta, \phi)&=&\int d\tau \int {d^3 \xv} \left[
  i\frac{\Spin}{a^3}\dot{\phi}(1-\cos\theta)+
   \frac{J\Spin^2}{2}\left( (\nabla\theta)^2+\sin^2\theta(\nabla\phi)^2
\right)
\right.
\nonumber\\
&&-\left.\frac{K}{2}\Spin^2\cos^2\theta +\frac{K_{\perp}}{2}
 \Spin^2\sin^2 \theta\cos^2\phi \right] +\Delta S_{\rm dis}
. \label{Smag}
\eeqa
The spin, $S=\Ms /2$, and the exchange coupling $J$[J/m] are determined
microscopically by Eqs.(\ref{MF}) and (\ref{Jdef}).

Without dissipation, $\Delta S_{\rm dis}=0$, the action Eq.(\ref{Smag}) has
a classical solution of a planar domain wall, where the spin configuration
changes only in one direction, which we choose as $x$-axis, as depicted in
Fig.\ref{FIGDW}(a). The position of the wall is  at $x=Q$.
Because of the anisotropy, the spin points to the $\pm$z direction at
infinity and it rotates
in the $yz$-plane (due to transverse anitsotropy energy that unfavors the
$x$-direction) around $x=Q$.
For a wall moving with a small velocity $\dot{Q} \ll c$, where
$c\equiv K_\perp \lambda Sa^3/\hbar$, the classical solution is given by
\beq
\cos\theta(\xv, \tau) = \tanh \frac{x-Q(\tau)}{\lambda}, \;\;\;\;\;\;\;\;
\cos\phi(\xv,\tau)\simeq i\frac{\dot{Q}}{c} \ll 1 .\label{DWT}
\eeq
Here the width of the wall, $\lambda$, is given by
\beq
\lambda\equiv \sqrt{\frac{J}{K}} .
\eeq

The dynamics of the wall coordinate $Q(\tau)$ is determined as follows.
For the configuration of Eq.(\ref{DWT}), the Lagrangean in Eq.(\ref{Smag})
(terms in square bracket) is written in terms of a collective coordinate
$Q$ as (adding irrelavant constant)
$L_w= -({1}/{2})M_{\rm w}\dot Q^2$
where the mass of the wall is given by
\beq
M_{\rm w}\equiv\frac{2\Awall}{K_\perp\lambda},    \label{Mwall}
\eeq
$\Awall$ being the area of the wall.
The strength of the pinning potential for $Q$ produced by the defect
is to be related to the coercive field $H_c$ of the system.
The magnetic field $H$ produces a linear potential $\propto HQ$ and thus
makes the pinning center $Q=0$ metastable (see Fig. \ref{FIGDWpot}(b)).
In the experimental situations, the field is set very close to the
coercive field $H_c$,
where the energy barrier vanishes,  for the tunneling rate to be large enough.
In this small energy barrier case, the total potential for $Q$ produced
by the defect and magnetic field is well approximated
by a sum of  harmonic and cubic potential\cite{SCB}.
\beq
V(Q)=\half M_{\rm w} \omega_0^2 Q^2\left(1-\frac{Q}{Q_0}\right). \label{pot}
\eeq
Here the attempt frequency $\omega_0$ and the tunneling barrier width $Q_0$
is shown to be
expressed in terms of macroscopic parameters in the leading order of
$\epsilon\equiv(H_c-H)/H_c \ll 1$ as\cite{SCB,magst}
\beqa
\omega_0 &\simeq& \mu_0 \frac{(\hbar\gamma)^2 S}{a^3}\sqrt{h_c}
\epsilon^{\frac{1}{4}} \simeq
 10^{11}\times \sqrt{h_c}\epsilon^{\frac{1}{4}} \,\,\, ({\rm Hz})
 \simeq 5\times \sqrt{h_c}\epsilon^{\frac{1}{4}} \,\,\, ({\rm K})
\label{omegazero}\\
Q_0 &=& \sqrt{\frac{3}{2}}\sqrt{\epsilon}\lambda ,\label{Qzero}
\eeqa
where $h_c\equiv H_c/(2\hbar\gamma S /a^3)$ is the ratio of the coercive
field to the magnetic moment per unit volume.
The typical values of this parameter is $h_c\simeq 10^{-4}\sim10^{-2}$.
Note that the barrier width $Q_0$ is much smaller than the domain wall
width $\lambda$ for the case
of shallow potential ({\it i.e.,} small $\epsilon$).
The barrier height is proportional to the number of spins in the wall
$N\equiv\Awall\lambda/a^3$, and is expressed as
$
U_{\rm H}\simeq \mu_0 ({(\hbar\gamma S)^2 }/{a^3})N h_c
\epsilon^{\frac{3}{2}} $$\simeq
 5\times N{h_c}\epsilon^{\frac{3}{2}} \,\,\, ({\rm K})$.

{}From these considerations, the dynamics of $Q$ is determined by the
Lagrangean
\beq
L(Q)=  \half M_{\rm w}\dot{Q}^2
  +  \half M_{\rm w} \omega_0^2 Q^2\left(1-\frac{Q}{Q_0}\right) .
\eeq
The tunneling rate of the variable $Q$ from the metastable state $Q=0$ of
the potential $V(Q)$ is estimated
using the bounce solution of $L(Q)$, which describe $Q$ travelling from $0$
at $\tau=-\infty$
to $Q_0$ at $\tau=0$ and goes back to $Q=0$ at $\tau=\infty$\cite{CC}.
It is explicitly given as
\beq
Q(\tau)=Q_0\frac{1}{\cosh^2\frac{\omega_0\tau}{2}}. \label{bounce}
\eeq
The exponent of the tunneling rate without dissipation is given by the
value of action
$B\equiv \int d\tau L(Q)$ estimated along the bounce solution.
Taking into account the prefactor, the tunneling rate without dissipation is
estimated as\cite{Sta}
\beq
\Gamma_0=Ae^{-B} , \label{rate0}
\eeq
where
$
A\simeq 10^{11} h_c^{\frac{3}{4}}N^{\half}\epsilon^{\frac{7}{8}}
\,\,\,({\rm Hz})
$ and $
B\simeq h_c^\half N \epsilon^{\frac{5}{4}}\cite{exponent}.
$
Because $h_c$ and $\epsilon$ are small, MQT of a domain wall is observable
even for very large number of spins if the dissipation is negligible.
For example, for $h_c=10^{-4}$ and $N=10^{6}$, MQT is observable
({\it e.g.,} $\Gamma_0 \gsim 10^{-4}$Hz) by
setting the magnetic field within 1\% of the coercive field ({\it i.e.},
$\epsilon\lsim10^{-2}$)(see dashed line in Fig.\ref{FIGgamma}).
Instead, $\epsilon\lsim2\times10^{-4}$ is needed for $N=10^8$.

The crossover temperature $T_{\rm co}$ from the thermal activation to the
quantum tunneling is estimated by the relationship $B=U_{\rm H}/T_{\rm co}$
where $U_{\rm H}$ is the barrier height of the potential.
For the present problem, the crossover temperature is roughly the same as the
attempt frequency $\omega_0$;
\beq
T_{\rm co}\simeq 5\times \sqrt{h_c} \epsilon^{\frac{1}{4}}\,\,\,({\rm K}).
\label{Tc0}
\eeq
%

\subsection{Effect of dissipation}
Now we discuss the effect of dissipation due to itinerant electrons,
$\Delta S_{\rm dis}$.
Within the calculation based on the bounce solution, the tunneling rate
Eq.(\ref{rate0}) is reduced by dissipation to be\cite{CL}
\beq
\Gamma=Ae^{-(B+\Delta {S}_{\rm dis})} =\Gamma_0 e^{-\Delta {S}_{\rm dis}} .
\label{rate}
\eeq
Here, to the lowest order in dissipation, the exponent
$\Delta {S}_{\rm dis}$ is evaluated as the value of the action
Eq.(\ref{Sdis})
 estimated along the domain wall solution Eq.(\ref{DWT}) with
$Q(\tau)$ given by the bounce Eq.(\ref{bounce}).
For the present planar domain wall solution Eq.(\ref{DWT}),
 $\Theta_\qv$ is obtained as
\beq
\Theta_\qv(\tau)= -\pi A_w e^{-iq_1 Q(\tau)}\frac{1}{\cosh
\frac{\pi}{2}\lambda q_1}
\delta_{q_2,0}\delta_{q_3,0}, \label{thetbounce}
\eeq
where $q_i$'s are the $i$-th component of the momentum $\qv$,
and $\Delta {S}_{\rm dis}$ reads as
\beq
\Delta S_{\rm dis}\simeq N\frac{(\kfu^2-\kfd^2)^2a^4}{32\pi\lambda a}
\int d\tau\int d\tau' \frac{(Q(\tau)-Q(\tau'))^2}{(\tau-\tau')^2}
 \int^{(\kfu+\kfd)\frac{\pi}{2}\lambda}_{(\kfu-\kfd)\frac{\pi}{2}
\lambda}
\frac{dx}{x} \frac{1}{\cosh^2 x}  , \label{DS3}
\eeq
where we have used $\epsilon\ll 1$ and the fact that the integral is dominated
by small momentum transfer; $x \lsim1$.
In evaluating the time integral, we note that the bounce solution
$Q(\tau)$ is localized within $|\tau|\lsim \omega_0^{-1}$, and
we approximate the function $(Q(\tau)-Q(\tau'))$ as\cite{cut}
\beqa
Q(\tau)-Q(\tau')=\left\{ \begin{array}{llr} -Q_0 & & |\tau|\geq 2\omega_0^{-1},
|\tau'|\leq \omega_0^{-1} \\
                                             Q_0 & & |\tau|\leq \omega_0^{-1},
|\tau'|\geq 2 \omega_0^{-1}
  \end{array}\right. .\label{Zap}
\eeqa
The action then results in
\beq
\Delta  S_{\rm dis}\equiv \eta N\epsilon .\label{DSf}
\eeq
The strength of dissipation is given by
\beq
\eta= \frac{3\ln 3}{16\pi}{(\kfu^2-\kfd^2)^2 a^4}\frac{\lambda}{ a}
\int^{(\kfu+\kfd)\frac{\pi}{2}\lambda}_{(\kfu-\kfd)\frac{\pi}{2}\lambda}
dx\frac{1}{x} \frac{1}{\cosh^2 x},
\label{etadef}
\eeq
and its asymptotic behaviors are
\beqa
\eta(\lambda)\simeq  \frac{ 3\ln3}{16\pi}(\kfu^2-\kfd^2)^2 a^4 \times
  \left\{ \begin{array}{cr}
      \frac{\lambda}{a}\ln\frac{\kfu+\kfd}{\kfu-\kfd} & \,\,\,
 \lambda(\kfu+\kfd)\ll 1 \\
     \frac{4}{\pi}\frac{1}{(\kfu-\kfd) a}
           e^{-\pi\lambda(\kfu-\kfd)}         &\,\,\, \lambda(\kfu-\kfd)
\gg 1
 \end{array} \right.  \label{Ilambda}
\eeqa
Thus the effect of dissipation is exponentially small for a thick wall
$\lambda(\kfu-\kfd)\gg 1$.
In the case of an bulk iron, for example, the domain wall thickness is about
$\lambda\simeq200 $\AA \ and so the effect of dissipation
from the itinerant electron would be negligible.
For a thin wall, on the other hand, as realized in {\it e.g.,} SmCo$_{5}$
($\lambda\simeq12$\AA), $\eta$ may be large.
In Fig. \ref{FIGeta}, $\eta$ is plotted as a
function of $\lambda/a$ for $\ktil\equiv { (\kfu+\kfd)a}/{2}=3$ and
$\delta\equiv{ (\kfu-\kfd)a}/{2\ktil}=0.05,0.1$ and $0.2$.
It is seen that $\eta$ can be of the order of 0.1 for $\lambda/a\simeq
2$ and $ \delta\simeq 0.05$.
In contrast to the present metallic case, the Ohmic dissipation vanishes
in insulators at absolute zero, since in that case,
the major source of dissipation is the magnon, which has a
gap due to the anisotropy.

\begin{figure}[bt]
\vspace{3.6cm}

\caption{  (left)
  The strength of the dissipation $\eta$ given by Eq.(\protect\ref{etadef})
as a function of the width of the wall $(\lambda/a)$,
for $\delta\equiv (\kfu-\kfd)a/(2\tilde k_0)= 0.05, 0.1$ and $0.2$ with
$\tilde k_0 \equiv(\kfu+\kfd)a/2 =3.0$.
  \label{FIGeta} }

\caption{  (right)
  The tunneling rate $\Gamma$ (solid line) for a typical case of an metal
$(\eta=0.1)$ with total number of spins in the wall
$N=10^4$ and $10^6$, respectively. The dashed lines are the rate $\Gamma_0$
in the absence of dissipation.
The parameter is taken as $h_c=10^{-4}$.
  \label{FIGgamma} }

\end{figure}

In the presence of dissipation, the tunneling rate $\Gamma$ is given by
Eq.(\ref{rate})  with the action given by Eq.(\ref{DSf}).
For a typical case of metal; $ \eta=0.1$, $\Gamma$ is shown in  Fig.
\ref{FIGgamma} for $N=10^4$ and $10^6$ with $h_c=10^{-4}$ .
The value $N=10^4$ corresponds, for instance, to a material with the wall
thichness of $\sim10$\AA\ and the area of $200$\AA$\times 200$\AA.
The rate $\Gamma_0$ without dissipation defined by Eq.(\ref{rate0}),
 which corresponds to the case of an insulator are also
plotted by dashed lines.
It is seen that to obtain an observable tunneling rate ($\gsim10^{-5}$Hz) in
metals, smaller value of $\epsilon$ by a factor of about $10^{-2}$ is needed
than in insulators.
%

\section{Dissipation in Multi-band Model}
\label{SECsd}
Our preceeding calculation shows strong dissipative effects for a thin wall
in weak ferromagnet.
In convensional bulk metals, these two conditions might not be easy to be
satisfied simultaneously.
However, in realistic situations, the existence of multi-band ({\it e.g.,}
$s$-$d$ two band model) must be taken into
account.

We consider the simplest case of the multi-band,
the $s$-$d$ model, where the localized magnetic moment is due to $d$
electron and the current is carried by $s$ electron.
Due to the $s$-$d$ mixing, the antiferromagnetic exchange
interaction between localized moment $\Mv(\xv)$ of $d$ electrons and the $s$
electron is present\cite{SW}
\beq
H_{sd}=g \sum_\xv \Mv(\xv) \cdot (c^{(s)\dagger} {\sigbf} c^{(s)})_\xv .
\label{sd}
\eeq
Hence the Lagrangean of the $s$ electron is of the same form as Eq.(\ref{L1}),
except that the last
term in Eq.(\ref{L1}) is absent here, since the
Coulomb interaction among $s$ electron is neglected.
The calculation is carried out in the same way as in \S \ref{SECeff} and
\ref{SECdw}.
Similarly to Eq.(\ref{etadef}), the strength of dissipation due to $s$
elctron
is obtained as\cite{sdlocal}
 $(\Delta S_{\rm dis}^{(s)}\equiv \eta^{(s)} N\epsilon)$
\beq
\eta^{(s)}=\frac{3\ln 3}{64\pi}{[(\kfu^{(s)})^2-(\kfd^{(s)})^2]^2 a^4}
\frac{\lambda}{ a}
\int^{(\kfu^{(s)}+\kfd^{(s)})\frac{\pi}{2}\lambda}
_{(\kfu^{(s)}-\kfd^{(s)})\frac{\pi}{2}\lambda}
dx\frac{1}{x} \frac{1}{\cosh^2 x},\label{etas}
\eeq
where $\kfs^{(s)}$ is the fermi momenta of the $s$ electron.
This expression is the same form as $\eta$ (Eq.(\ref{etadef}))\cite{noRPA},
and hence
the behavior of $\eta^{(s)}$ is accordingly read from Fig.\ref{FIGeta}, and
 $\eta^{(s)}$  can be $O(0.1)$
if the lower bound of the integration, $(\kfu^{(s)}-\kfd^{(s)})\pi\lambda/2$,
is smaller than unity.
Hence if the $s$-$d$ coupling is sufficiently small: $(\lambda\kf^{(s)}\alpha
\ll 1)$
$(\alpha\equiv (g \Ms/\ef^{(s)}))$,
large value of $\eta^{(s)}\simeq0.1$ is possible even in strong ferromagnet.
The expression of $\eta^{(s)}$ in this limit is
\beq
\eta^{(s)}\simeq \frac{3\ln 3}{16\pi}(\kf^{(s)} a)^4 \left(\frac{\lambda}{a}
\right) \alpha^2\left|\ln\left(\frac{\pi}{2}\lambda\kf^{(s)}\alpha\right
)\right|.
\eeq
For $\alpha=0.05$, $(\lambda/a)\simeq2$ and $\kf^{(s)} a=3$, for example,
$\eta^{(s)}\simeq 0.02$.
The result indicates that in the multi-band systems,
dissipation due to Stoner excitation will be significant for a thin wall
even in the case of strong ferromagnet.

\section{Dissipation due to Eddy Current}
\label{SECeddy}
Besides the direct coupling to the electron spin current, the moving wall in
metals also interacts with the charge current (eddy current) via induced
 electric field governed by Faraday's law as noted by Chudnovsky
{\it et al.}\cite{CIS}.
The dissipation due to this eddy current is Ohmic, but this effect turns
out to be small for a case of meso- or microscopic wall, as discussed below.

The electric field induced by the change of the magnetization is
calculated
 from the Maxwell equation
\beq
\nabla\times{\bf E}=-\mu_0\dot{\bf M},
\eeq
where ${\bf M}\equiv (2\hbar\gamma S/a^3)\nv$ is the magnetization
vector.
For the case of a moving domain wall, where the angles are given by
Eq.(\ref{DWT}), only the $y$-component $ E_y$ is important, which is obtained
as
\beqa
E_y(\xv,\tau)\simeq\mu_0\left(\frac{2\hbar\gamma S}{a^3}\right)
\dot Q(\tau) \times \left\{ \begin{array}{lr}
  \tanh\frac{x-Q}{\lambda} & |x-Q| \lsim L  \\
  \frac{1}{2\pi}\frac{L^2}{(x-Q)^2} {\rm sgn}(x-Q) & |x-Q| \gg L \end{array}
  \right., \label{electric}
\eeqa
where $L$ is the linear dimension of the crosssection of the wall
($A_w=L^2$).
For $L\gg \lambda$, the electric field is almost constant over the distance
$|x-Q|\lsim L$ and decays for a larger distance.

The electromagnetic coupling to electrons is
\beq
H_{\rm EM}=e\sum_\xv {\bf A}\cdot{\bf j},
\eeq
where the electron current is given by
$
{\bf j}(\xv,\tau)=-({i}/{2m})(c^\dagger\nabla c-\nabla c^\dagger c),
$
and the vector potential is related to the electric field by
${\bf E}=-\dot{\bf A}$.
Since ${\bf E}$ and hence ${\bf A}$ depend on $Q$, this coupling
$H_{\rm EM}$
gives rise to dissipation effects on the motion of $Q$.

Treating $H_{\rm EM}$ perturbatively to the second order and integrating
out the electron, we obtain the following term in the effective action for
$Q$;
\beq
\Delta S_{\rm ch}=-\frac{1}{2}\int d\tau\int d\tau' \frac{1}{\Vtil}\sum_\qv
 A_\mu(\qv,\tau)A_\nu(-\qv,\tau')<j_\mu(\qv,\tau)j_\nu(-\qv,\tau')>,
\eeq
where $A_\mu(\qv,\tau)$ and $j_\mu(\qv,\tau)$ are the Fourier transforms
of the vector potential and the current.
{}From the behavior of $ E_y(\xv,\tau)$, it is easily seen that $A_y(\qv,\tau)$
is well approximated as
\beq
A_y(\qv,\tau)\simeq
 \mu_0\left( \frac{2\hbar \gamma S}{a^3}\right)\frac{1}{iq_x}e^{iq_x Q}
A_w L \times \delta_{q_y,0}\delta_{q_z,0}
\eeq
for $q_x\lsim L^{-1}$ and $A_y\simeq 0$ for $q_x\gg L^{-1}$.

We consider the disordered case where $L$ is much larger than the mean
free path of electron $\ell$;
\beq
L\gg \ell,
\eeq
since such a case will be of interest in experimental situations.
Because of this condition, the correlation function
$<j_\mu(\qv,\tau)j_\nu(-\qv,\tau')>$ can be approximated by its value at
$q=0$.
In this case, the Ohmic contribution is written in terms of the conductivity
$\sigma$ as
\beq
<j_\mu(\qv,\tau)j_\nu(-\qv,\tau')>\simeq \frac{\sigma}{\pi}\int_0^\infty
\omega d\omega
e^{-\omega|\tau-\tau'|}  =\frac{\sigma}{\pi(\tau-\tau')^2}.
\eeq
The action is thus written as
\beq
\Delta S_{\rm ch}\simeq \frac{1}{4\pi}\left[\mu_0\left(\frac{2\hbar\gamma S}
{a^3}\right)\right]^2 \sigma A_w L\times \int d\tau\int d\tau'
\frac{(Q(\tau)-Q(\tau'))^2}{(\tau-\tau')^2},
\eeq
where we have used $(Q(\tau)-Q(\tau'))/L\ll 1$
and subtracted an irrelevant constant.
Therefore the strength of Ohmic dissipation $\eta^{\rm (ch)}$ due to the
eddy current, defined by $\Delta S_{\rm ch}=\eta^{\rm (ch)}N\epsilon$
($N=A_w\lambda/a^3$ and $\epsilon$ arises
from $Q_0\simeq \sqrt\epsilon \lambda$), is\cite{eddy}
\beq
\eta^{\rm (ch)}\simeq\frac{1}{4\pi}\left[\mu_0\left(
\frac{2\hbar\gamma S}{a^3}
\right)\right]^2 \frac{a^5}{\hbar} \sigma \left(\frac{\lambda}{a}\right)
\left(\frac{L}{a}\right).
\label{etach}
\eeq
For a typical metal, $\sigma\simeq 10^8$[$\Omega^{-1}$m$^{-1}$],
it is  $\eta^{\rm (ch)}\simeq 10^{-7}\times ({\lambda}/a)({L}/{a})$
if we choose $S\simeq1$ and $a=3\times 10^{-10}$[m].
Hence, as far as $(\lambda/a)\lsim 10^2$, which is commonly satisfied,
dissipation due to eddy current
is small in mesoscopic wall we are interested in; $(L/a)\lsim 10^4$.

\section{Discussions and Conclusion}
\label{SECdisc}
We have studied the case of a domain wall where the plane of the wall is
parallel to the easy ($yz$-)plane.
It is also possible to consider based on the action Eq.(\ref{Smag})
another configuration where spin direction changes in the easy ($z$-)
direction of spin as shown in Fig. \ref{FIGDW}(b)\cite{dwb}.
For the MQT of this wall, the results obtained above do not change.

A contribution from the magnon pole in the correlation function
$<\Jv\Jv>$ leads to the
dissipation with a gap $\Delta_0$ due to anisotropy.
In the present problem of tunneling, the effect needs to be evaluated
quantitatively\cite{TF,super}.
The strength of dissipation $\eta^{\rm (pole)}$, which is
defined by the value of the action devided by $(N\epsilon)$, is obtained as
$
\eta^{\rm (pole)} =({3\pi}/{40})\Ms \exp({-({\Delta_0}/{\omega_0}})).
$
Since experiments are usually carried out in highly anisotropic materials
with
 $\Delta_0/\omega_0\simeq 10$, this contribution is very small
compared to the Ohmic dissipation for the case of a thin wall.

Our result shows a distinct difference between MQT of thin
walls in metallic and insulating magnets.
Unfortunately the experiments carried out so far appear not yet be able to
test the theoretical prediction of the effect of dissipation due to
itinerant electrons.
The experimental result on small ferromagnetic particles of
Tb$_{0.5}$Ce$_{0.5}$Fe$_2$ suggests the motion of a domain wall via MQT
below $T_{co} \simeq 0.6$K\cite{PSB}.
In this experiment, the width of the domain wall is about $30$\AA\ and
according to our result, $\eta\propto \exp[-\pi\lambda(k_{{\rm F}\uparrow}
-k_{{\rm F}\downarrow})]$,
the dissipation from electron spin current is negligible for such a thick
wall.
This may be the reason why the result of the crossover temperature
$\sim0.6$K is roughly in agreement  with the theory\cite{Sta}
  without dissipation
 (Eq.(\ref{Tc0}) with $h_c \sim 4\times 10^{-4}$ and $\epsilon \sim0.1$).
On the other hand, the domain wall in SmCo$_5$ is
very thin $\lambda\simeq12$\AA, and our
result suggests strong effect of dissipation due to Stoner excitation,
which will be interesting to observe.
Experiments on bulk crystal of
SmCo$_{3.5}$Cu$_{1.5}$ with very thin walls (a few times $a$) have
been performed\cite{UB}, although quantitative comparison
is not easy since many walls will participate in the relaxation processes of
the magnetization in these experiments.

\label{SECconc}
To conclude, we have studied the macroscopic quantum tunneling of a
domain wall in a metallic ferromagnet based on the Hubbard model.
The effective action of the magnetization is obtained to be the same form
as a ferromagnetic Heisenberg model but with a non-local (retarded) part, that
 describes the dissipative effect on magnetization.
In contrast to the case of insulators,
the Ohmic dissipation exists even at zero temperature
because of the gapless Stoner excitation in the itinerant electron.
The effect is negligible for a thick domain wall where experiments so far
have been carried out.
On the other hand, important effects of the Ohmic dissipation are expected
in thin domain walls with thickness $\lambda$ comparable
to the inverse of the difference of the fermi momenta $(\kfu-\kfd)^{-1}$.
In the strong ferromagnets, the Ohmic dissipation arising from the magnetic
band is weakend. In this case the existence of multi-band needs to be taken
into account whose effect can be significant
even in strong ferromagnets.
Dissipation due to the eddy current has been shown to be negligibly small
for a mesoscopic system.
We believe the observation of the dissipation due to the Stoner excitation
is within the present experimental attainability.

In the case of a single domain (ferro-)magnets, where another typical MQT
can occur\cite{ES}, our analysis
implies that the dissipation due to the conduction electron is irrelevant
 since only the weak component of the electron excitation
({\it i.e.,} $q=0$) is coupled to the uniform magnetization. This fact is
seen as a vanishing $\eta$ (Eq. (\ref{Ilambda})) in the limit of single domain
magnet, $\lambda\rightarrow \infty$.

We have investigated the MQT in metal paying attention to the role of the
conduction electron.
On the other hand the electron transport properties should be affected by the
MQP of magnetism.
This new possibility will make the qunatum tunneling in magnetic system
more exciting in metals.
In fact the experimental study in this direction has been started
recently\cite{Hong},
where the resistivity of a mesoscopic Ni wire was measured in the magnetic
field.
Small jumps have been observed there, which may be related to the MQT of domain
walls at low temperature.
The novel feature of this experiment is that it is
on a {\it single} mesoscopic sample, hence one will be able to detect a {\it
single}
MQP event.
This is in contrast to the magnetization measurement
using SQUID where in general a large number of small particles are needed to
get a sufficient signal, and hence one must worry about the distribution of
energy barriers in its interpretation.
Another new possiblity in the study of the mutual effects between the
conduction
electron and the MQP of magnetism would be magnetic dots in
semiconductors\cite{Awa},
where one can control the interaction by changing the electron density.

The MQP in magnetic metals and semiconductors will be a very important field
also
from the point of view of application and there are, we believe, a lot of work
to be done in this new field.

\acknowledgements
%

This work is financially supported by Ministry-Industry Joint Research
program "Mesoscopic Electronics" and by Grant-in-Aid for Scientific Research
on Priority Area, "Electron Wave Interference Effects in Mesoscopic
 Structure"
(04221101) and by Monbusho International Scientific Research Program:Joint
Research "Theoretical Studies on Strongly Correlated Electron Systems"
(05044037) from the Ministry of Education, Science and Culture of Japan.
G. T. is supported by the Special Grant for Promotion of Research from the
Institute of Physical and Chemical Research (RIKEN).

%

%
%

\begin{thebibliography}{}
\bibitem{dwb}
This configuration is possible if the anisotropy energy along the easy axis
is much larger than the demagnetization energy.
\bibitem{Sta}
 P. C. E. Stamp, Phys. Rev. Lett. {\bf 66}, 2802 (1991).
\bibitem{CIS}
 E. M. Chudnovsky, O. Iglesias and P. C. E. Stamp, Phys. Rev.
{\bf B46}, 5392 (1992).
\bibitem{SCB}
 P. C. E. Stamp, E. M. Chudnovsky and B. Barbara, Int. J. Mod.
Phys. {\bf B6}, 1355 (1992).
\bibitem{CL} A. O. Caldeira and A. J. Leggett, Phys. Rev. Lett. {\bf 46},
211 (1981);
 A. O. Caldeira and A. J. Leggett, Ann. Phys. {\bf 149}, 374 (1983).
\bibitem{GK}
 A. Garg and G-H. Kim, Phys. Rev. Lett. {\bf 63}, 2512 (1989);
 A. Garg and G-H. Kim, Phys. Rev.  {\bf B43}, 712 (1991);
 H. Simanjuntak, J. Low Temp. Phys. {\bf 90}, 405 (1992).
\bibitem{Gargnuc}
 A. Garg, Phys. Rev. Lett. {\bf 70}, 1541 (1993).
\bibitem{UB}
 M. Uehara, B. Barbara, B. Dieny and P. C. E. Stamp, Phys. Lett.
{\bf A114}, 23 (1986);
M. Uehara and B. Barbara, J. Physique, {\bf 47}, 235 (1986).
\bibitem{PSB}
 C. Paulsen, L. C. Sampaio, B. Barbara, D. Fruchard, A.
Marchand, J. L.  Tholence and M. Uehara, Phys. Lett. {\bf A161}, 319 (1991);
C. Paulsen, L. C. Sampaio, B. Barbara, R. T-Tachoueres, D. Fruchart, A.
Marchand, J. L. Tholence and M. Uehara, Europhys. Lett. {\bf 19}, 643 (1992).

\bibitem{TF}
 G. Tatara and H. Fukuyama, Phys. Rev. Lett. {\bf 72}, 772 (1994);
J. Phys. Soc. Jpn. {\bf 63}, 2538 (1994).
\bibitem{continum}
The results would be
valid even for a thin domain wall with width of a few times lattice
constant.
\bibitem{magfield}
We neglect the effect of magnetic field on electronic states.
This is justified as long as $UM\gg\gamma H$.
In experimental situations with the magnetic field of $\lsim1$T and
$U\simeq 10$eV,
this condition reduces to $M\gsim 10^{-4}$ in unit of the Bohr magneton,
which is easy to satisfy.

\bibitem{Pra}
 V. Korenman, J. L. Murray and R. E. Prange, Phys. Rev.
{\bf B16}, 4032 (1977).
\bibitem{meanfield}
These are the results of the Hartree-Fock theory, which describes
the essential features of itinerant ferromagnetism but should not be taken
literally in comparison with the actual experiments.
\bibitem{RPA}
RPA summation is needed since we have not taken into account in the
Coulomb interaction the process with finite momentum transfer,
$
\Huf\equiv -({U}/{2 \Vtil })\sum_{\qv\neq0}\sum_{\kv\kv'}
(a^\dagger_{\kv+\qv} \sigma_z a_\kv )(a^\dagger_{\kv'-\qv} \sigma_z a_{\kv'}
),$
in the determination of the magnitude of
the magnetization, which has been assumed to be uniform.
\bibitem{RS}
 C. Herring, in {\it Magnetism}, edited by G. T. Rado and H. Suhl
(Academic, New York,1966), Vol. IV.
\bibitem{ATF}
 K. Awaka, G. Tatara and H. Fukuyama, Jour. Phys. Soc. Jpn.,
{\bf 62}, 1939 (1993).

\bibitem{superohmic}
The super-Ohmic contributions, which are of higher orders of
$(\omega_0/\epsilon_{\rm F})$ in Eq. (\ref{Im}), are
smaller than the Ohmic one by a factor of
$(\omega_0/\epsilon_{\rm F})^2 \ll 1$ and hence are negligible.

\bibitem{demag}
We consider the case of strong anisotropy; $K,K_{\perp}\gg K_d$,
where $K_d\equiv \mu_0(\hbar\gamma)^2/a^6$ is the magnetostatic
energy due to demagnetization
($\gamma=e/(2m)$ is the gyromagnetic ratio
and $\mu_0$ is the magnetic permeability of a free space).
The calculation also applies to a ferromagnet with uniaxial
anisotropy $-K$ in $z$-direction, if one replaces $K_{\perp}\rightarrow K_d$.
\bibitem{kperp}
In the absence of the transverse anisotropy, $K_{\perp}$,
the domain wall cannot tunnel, since without this term $S_z$ is
conserved at each site.
This fact is expressed in Eq.(\protect\ref{Mwall}) as the divergence of
the domain wall mass as $K_\perp\rightarrow 0$.
\bibitem{aniel}
The effect of the anisotropy energy on the dissipation due to the itinerant
electron is neglected, since the correction would be small by the order of
$(Ka^3/U\Ms)\simeq 10^{-3}$ (for $Ka^3\simeq10$K, $\Ms \gsim0.1$).
\bibitem{magst}
A factor of $10^{11}$ (Hz) in the expression of $\omega_0$
arises from the magnetostatic energy due to the magnetization
of $(2\hbar\gamma S/a^3)\simeq 10^6$[A/m] ({\it i.e.,} $S\simeq O(1)$).

\bibitem{CC}
 C. G. Callan, Jr. and S. Coleman, Phys. Rev. {\bf D16}, 1762 (1977).
\bibitem{exponent}
The factor of $\epsilon^{5/4}$ arises
from the barrier height and width in the small $\epsilon$ limit
as seen by the WKB approximation;
$\Gamma_0 \propto$(barrier height)$^{1/2}
\times Q_0\propto \epsilon^{3/4}\epsilon^{1/2}$.
\bibitem{cut}
The integration must be cut off at short time $\sim \omega_0^{-1}$, since
in the calculation of $\Delta S_{\rm dis}$, we have made use of the bounce
solution with zero energy, that is, we have neglected the excited states
of variable $Q$.
This approximation would be valid only
for small energy transfer $\omega\lsim \omega_0$ in the
current correlation function $<\Jv\Jv>$.
See  Yu. Kagan and N. V. Prokof'ev, Zh. Eksp. Teor. Fiz. {\bf 90}, 2176 (1986)
[Sov. Phys. JETP {\bf 63}, 1276 (1986)].
\bibitem{SW}
 J.R. Schrieffer and P. A. Wolff, Phys. Rev. {\bf 149}, 491 (1966).
\bibitem{sdlocal}
The contribution to the local part of the effective
action due to the $s$ electron renormalizes the magnitude of
$S$ and $J$,
but this renormalization can be understood as already included in the
values of these quantities.
\bibitem{noRPA}
The numerical factor of Eq. (\protect\ref{etas}) is (1/4) times that of
Eq. (\protect\ref{etadef}) because no RPA summation is needed in the $s$-$d$
model.
\bibitem{eddy}
A rough estimate of $\eta^{\rm (ch)}$ is as follows.
By the definition of $\eta^{\rm (ch)}(\equiv \Delta S_{\rm ch}/(N\epsilon))$,
we can write
\[
\Delta S_{\rm ch}= N\frac{\eta^{\rm (ch)}}{\lambda^2} \int d\tau\int d\tau'
\frac{(Q(\tau)-Q(\tau'))^2}{(\tau-\tau')^2}.
\]
Noting that the integrand is regarded as rate of energy dissipation due to
the Joule heat, we have the relation
\[
\frac{N}{V}\frac{\eta^{\rm (ch)}}{\lambda^2}\dot Q^2 \simeq \sigma {\bf E}^2,
\]
where $V=A_w L$.
By use of Eq.(\ref{electric}), we obtain the expression Eq.(\ref{etach})
correct up to a numerical factor.

\bibitem{super}This is because, in contrast to the case of quantum coherence
problem\protect\cite{ATF}, the Ohmic dissipative action is not divergent
at long time, and hence the contribution from the Ohmic
dissipation is not qualitatively distinct than those from the
super Ohmic one and dissipation processes with an excitation gap.
\bibitem{ES}
 M. Enz and R. Schilling, J. Phys. {\bf C 19}, 1765; L711 (1986);
 J. L. van Hemmen and A. S\"ut\H{o}, Europhys. Lett., {\bf 1},
481 (1986); Physica {\bf 141B}, 37 (1986);
 E. M. Chudnovsky and L. Gunther, Phys. Rev. Lett. {\bf 60}, 661
 (1988);
 B. Barbara and E. Chudnovsky, Phys. Lett. {\bf A145}, 205 (1990);
 I. V. Krive and O. B. Zaslavskii, J. Phys., {\bf 2}, 9457 (1990).

\bibitem{Hong}K. Hong and N. Giordano, in this volume; N. Giordano and J. D.
Monnier,
Physica {\bf B194-196}, 1009 (1994).
\bibitem{Awa}J. J. Baumberg, D. D. Awaschalom, N. Samarth, H. Luo and J.K.
Furdyna,
Phys. Rev. Lett. {\bf 72}, 717 (1994).
\end{thebibliography}
\end{document}